Comments on "Evidence for Nuclear Emissions During Acoustic Cavitation"; by R.P. Taleyarkhan et al..,Science **295**, 1868, March 8, 2002

S.J. Putterman (1), L.A. Crum (2); K. Suslick (3)
(1) Physics Department, University of California, Los Angeles CA 90095
(2)Applied Physics Laboratory, University of Washington, Seattle WA 98105
(3) Department of Chemistry, University of Illinois, Urbana, IL 61801

Abstract: In a paper recently published in Science, Taleyarkhan et al. claimed to observe fusion from acoustic cavitation and the associated phenomenon of sonoluminescence. Although, this is a worthwhile line of investigation we explain why, in our opinion, their data neither proves nor disproves this possibility.

**Sonoluminescence** is an amazing phenomenon. Under the action of a sound field whose pressure is only slightly higher than ambient [5] {Numbered references are keyed to Taleyarkhan et al.and given at the end of this paper} a bubble can undergo pulsations that locally focus the pressure to a kilobar or more, and emit picosecond flashes of light [8] whose spectrum indicates that the surface of the bubble is hotter than the surface of the sun [Vazquez et al. Optics Letters **26**,575,2001]. Is the interior of the bubble so hot and so dense that this tabletop apparatus could generate thermonuclear fusion when its contents are deuterated? Although this possibility was proposed theoretically some years ago [4] the first experiments to claim to observe such a phenomenon have very recently been reported. While we are sympathetic with the goals of Taleyarkhan et al. ["T et al."], we take this opportunity to explain why we found their paper unready for publication and why we still find their published results unconvincing.

**Tritium** is a major byproduct of D-D fusion and in the first manuscript version of Taleyarkhan et al. [dated June 2001 and referred to below as 'T1'] the authors stated that "…to search for tritium several smear samples were taken from the surface of the vacuum pump outlet line, the table on which the experiment was placed and the plastic walls of the enclosure around the cavitation chamber. Results in Table I are consistent with the presence of measurable amounts of tritium following the experiments with D-acetone". As reviewers, we interpreted this statement as not supportive of d-d fusion, but rather as indicating that the ORNL lab in which these experiments were undertaken was contaminated with tritium. Our conclusion should be contrasted with the claim in T1 that other than the cavitation-generated tritium found on the smears samples, "No other potential source of tritium is known to exist in the experimental laboratory [T1]." The origin of this contamination (in our opinion), has not been revealed, or even discussed. The effect of such contamination on the data reported in Science and why these tritium measurements were originally interpreted as signal remains unexplained.

The raw data for two examples of **neutron coincidences with sonoluminescence** [SL] were displayed in Figure 3 of T1. It was pointed out by the reviewers that these examples showed inconsistent time delays "$\Delta t_{SL-n}$" between SL and n. Although the flash of light and the [claimed] neutron from fusion are emitted simultaneously at the moment of implosion, the value of $\Delta t_{SL-n}$ is not zero due to the difference in the speed of light and a 2.5MeV neutron, as well as time delays introduced by the various detectors.

However, the value of $\Delta t_{SL-n}$ should be the same to within a few nanoseconds for all interesting events, i.e. those events where a bubble emits light and a 2.5MeV neutron from D-D fusion. Yet for the raw data displayed in T1 the values of $\Delta t_{SL-n}$ differed by more than 1 microsecond. As we noted, this time window is substantially greater than the accuracy with which the arrival of SL and n can, and should have been resolved. Thus at least one of these claimed signals is in fact noise. Apparently, the authors now make the claim that it must be possible to have nuclear emission from a bubble whose sonoluminescence is undetected. It is our position that the only true coincidences should be those with time delays in nanoseconds, not microseconds.

An example of a ***microphone* coincidences with SL and n** is shown in Figure 5A [T et al.] where it is noted that in "…mode 2 spurious signals were easily accommodated by taking data with and without cavitation and then subtracting coincidence data without cavitation from those taken with cavitation". Our concern here is that there can be no spurious signals without cavitation since the bubble does not launch a shock wave and create a microphone signal unless the sound field is on. The observation of background events (in this mode of data acquisition) by T et al. indicates that the actual experiment differs from their description.

The **magnitude** of the reported effect is highly suspicious, although we note that such a large effect is not ruled out by the first principles of physics. Only 1,000 neutrons, from the generator, pass through the acoustically active region per second, yet 50,000 neutrons/second and 700,000 tritons/second are claimed to be created due to cavitation fusion. This claimed yield is enormous and consequently its documentation should be obvious and without statistical ambiguity. If the energy and time spectrum of arrival of the radiation and SL were adequately resolved, say within a few tens of nanoseconds, the signal/noise would have been huge, since the effects of the pulsed neutron generator and other sources of background would have been largely removed. In fact, using an external source of neutrons as the seed is an undesirable way to look for neutrons from new sources of fusion. Although it is true that neutrons make **small cavitation nuclei**, this fact is not as important as claimed, because by the time the bubble collapses its interior will have lost almost all knowledge of its initial condition. In addition, although there may be small amounts of non-condensable gas in the liquid, vapor can also provide stiffness on bubble collapse.

**Calibration of the neutron detector** by use of Cs 137 and 14MeV neutrons from the pulsed neutron generator indicate an inconsistency in the experiment of T. et al. The maximum light generated by the interaction of a 14MeV neutron with their detector lies between channels 100-110. According to the light tables of N. P. Hawkes et al.[Nucl. Instr. Meth. Phys. Res. **A 476**, 190, (2002)] the edge for 2.5MeV neutrons should then lie in channel 10. The lower cutoff of channel 15-20 used by T. et al. would cut out the signal of interest. We also note that according to the light tables the edge for the 662KeV gamma [which creates a Compton edge electron of 478KeV whose light output is the same as a 1.25MeV proton] for Cs 137 should lie in channel 4 and not channel 30 as reported in T et al..

A positive aspect of T involves their use of a sound field [15atmospheres] which is higher than those recently used to study single bubble sonoluminescence [5]. This upscaling, {characteristic of multibubble sonoluminescence [11]} may have some advantages, though we note that the individual flashes of sonoluminescent light in this system involve less than a million photons and are therefore similar to SL in resonators running slightly above 1 atmosphere [4]. Perhaps this low yield is due to the non-spherical motion of neutron-seeded cavitation as was reported in 1969 [18].

Finally we take this opportunity to guess at some additional challenges that confront the experimental arrangement used by T et al.: 1) does crosstalk between the high voltage at the acoustic resonator affect the neutron generator or the photodetectors, and 2) do sonochemical byproducts (several of which are potent oxidants) affect the scintillation cocktail used to detect tritium, perhaps inducing false positives via chemiluminescence?

This line of investigation constitutes an excellent example of high-risk/high-gain research. Unfortunately, the possibility of a major discovery has been obscured by substandard experimental techniques.